\documentclass[journal]{IEEEtran}
\usepackage{amsfonts}
\IEEEoverridecommandlockouts

\ifCLASSINFOpdf
\else
\fi
\usepackage{epstopdf}
\usepackage{multirow}
\usepackage{cite}
\usepackage{stfloats}
\usepackage{color}
\usepackage{epsfig}
\usepackage{graphicx}
\usepackage{psfig}
\usepackage{epsf}
\usepackage{slashbox}
\usepackage{float}
\usepackage{amssymb }
\usepackage{cite}
\usepackage{stfloats}
\usepackage{color}
\usepackage{epsfig}
\usepackage{graphicx}
\usepackage{psfig}
\usepackage{epsf}
\usepackage{slashbox}
\usepackage{float}
\usepackage[cmex10]{amsmath}
\usepackage{algorithm}
\usepackage{algorithmic}
\begin{document}
\title{Energy-efficient  Resource Allocation for Wirelessly Powered  Backscatter Communications}
\author{\IEEEauthorblockN{Yinghui Ye, Liqin Shi, Rose Qingyang Hu, Guangyue Lu}\vspace*{-15pt}
\thanks{ Yinghui Ye (connectyyh@126.com) and   Guangyue Lu (tonylugy@163.com) are with the Shaanxi Key Laboratory of Information Communication Network and Security, Xi\rq an University of Posts \& Telecommunications, China. Yinghui Ye is also with the School of Telecommunications Engineering, Xidian University, China.}
 \thanks{Liqin Shi (email: liqinshi@hotmail.com) is with the School of Telecommunications Engineering, Xidian University, China.}
   \thanks{Rose Qingyang Hu is with the Department of Electrical and Computer
Engineering at Utah State University, U.S.A.
}
\thanks{This work was supported   by the scholarship from China Scholarship Council, the Natural Science Foundation of China (61801382), the   Science and Technology Innovation Team of Shaanxi Province for Broadband Wireless and Application (2017KCT-30-02),  the US National Science Foundations grants under the grants NeTS-1423348 and the EARS-1547312.}
}
\markboth{IEEE Communications Letters, No. XX, MONTH YY, YEAR 2019}
{Ye\MakeLowercase{\textit{et al.}}: Energy-efficient  Resource Allocation for Wirelessly Powered  Backscatter Communications}
\maketitle

\begin{abstract}
In this letter we consider a wireless-powered backscatter communication (WP-BackCom) network,  where the transmitter first  harvests energy from a dedicated energy RF source ($S$) in the sleep state. It subsequently backscatters information and harvests energy simultaneously through a reflection coefficient.
{\color{black}Our goal is  to maximize the achievable energy efficiency of the WP-BackCom network via jointly optimizing time allocation, reflection coefficient, and transmit power of $S$.}
The optimization problem is non-convex and challenging to solve. We  develop an  efficient Dinkelbach-based iterative algorithm to obtain  the optimal resource allocation scheme.  The study shows that for each iteration, the energy-efficient WP-BackCom network is equivalent to either the
network in which the transmitter always operates in the active state, or the network in which $S$  adopts the maximum allowed power. 
\end{abstract}

\IEEEpeerreviewmaketitle
\vspace{-5pt}
\section{Introduction}
\IEEEPARstart{T}{he}  limited battery lifetime of Internet of Things (IoT) devices can become a fundamental problem for massive  IoT  deployment. Several advanced technologies, e.g,  wireless-powered communication networks (WPCNs) \cite{6951347} and  backscatter communications  \cite{8368232}, have been  proposed to address this problem. One particular promising solution is to use the backscatter communication since it allows IoT devices to  modulate and reflect the incident RF signals instead of generating RF signals by itself and to harvest energy for circuit operation. Hence backscatter communications consume  much less energy than  WPCN involving oscillators, analog-to-digital/digital-to-analog converters \cite{8368232}. 

 In \cite{ kellogg2014wi}, the hardware  of backscatter communication prototypes was designed by leveraging  ambient  WiFi signals to realize backscatter and energy harvesting.
 In  \cite{7876867}, the authors   investigated the impacts of the time allocation and the reflection coefficient  on wirelessly powered backscatter communication system (WP-BackComs), where dedicated RF energy sources were deployed to power backscatter users. The coexistence  of harvest-then-transmit protocol  and backscatter communication was also investigated in   wireless-powered heterogeneous networks  \cite{7981380}, where  ambient RF signals and dedicated RF signals transmitted by a dedicated energy source  are considered.
 In addition to \cite{7876867, 7981380}, where the main focus was on system-level backscatter communications, the authors of \cite{8093703}   studied the joint design of time allocation and  reflection coefficient to maximize throughput in a typical backscatter communication scenario that consists of one RF energy source,  one backscatter user,  and one receiver. {\color{black} The outage probability  was also derived in a similar scenario \cite{8468064}}.
{\color{black} Backscatter communication has also been combined with other types of communication techniques, e.g., cognitive radio \cite{7937935, 8424210}, device-to-device \cite{8170328},  non-orthogonal multiple access \cite{8636518} and relaying  \cite{8490651,8302460}.} Other works on the  waveform design and detection algorithm design can  also be found in   \cite{8527670,7551180}.

\begin{figure}
  \centering
  \includegraphics[width=0.25\textwidth]{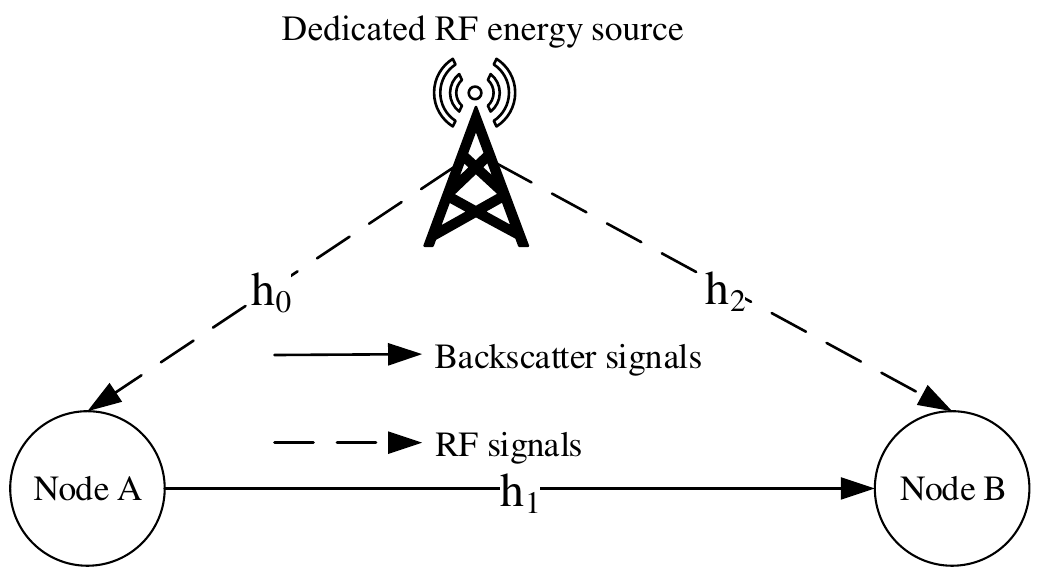}
  \caption{System Model}\label{1}
\vspace*{-18pt}
\end{figure}

   Although the aforementioned  works \cite{ kellogg2014wi,7551180,7876867,7981380,8093703,8468064,7937935, 8424210,8170328,8527670,8490651,8302460,8636518} have laid a solid foundation for understanding backscatter communications from various perspectives, e.g., hardware design and  {\color{black}spectral efficiency (SE)}, the energy efficiency (EE) of backscatter communication has not been studied yet. To the best of our knowledge, {\emph{this is the first work to consider energy-efficient resource allocation  problem in a  WP-BackCom network}}, where the  transmitter  modulates and reflects its information to the receiver  via RF signals from the dedicated RF energy source ($S$), as well as harvests energy to power its circuit.
  An efficient Dinkelbach-based iterative algorithm is developed to  determine the energy-efficient resource allocation scheme. {\color{black}Specifically,
   an optimization problem is  formulated to maximize the  EE  by jointly optimizing the time allocation, the reflection coefficient,  and the transmit power of $S$.}
   The  non-convex original optimization problem in the fractional form is  transformed into an equivalent  problem in the subtractive form based on  the  fractional programming. The transformed problem  can be cast into two convex EE maximization problems. In one problem the transmitter  operates in the active state and in the other problem $S$  adopts the maximum allowed power.

\section{System Model And Working Flow}
{\color{black}Fig. 1 shows a WP-BackCom network\footnote{{\color{black}BackCom can be classified into three major types: monostatic backscatter, bistatic backscatter, and ambient backscatter. The main differences among them have been summarized in \cite{8368232}.  Our considered system belongs to the bistatic backscatter due to the involved dedicated energy source.}} consisting of one $S$, one transmitter $A$  with backscatter circuits, and one receiver $B$.}
The nodes $B$ and $S$ have stable energy sources. Node $A$ is a battery-free node and backscatter communication is employed to realize information transfer and harvest energy for circuit operation. The harvested energy in each slot  is temporarily stored in a capacitor of node $A$.   {\color{black}For  simplicity, we assume that
  part of stored energy is used to power circuits and the rest is fully discharged in the same slot\footnote{{\color{black}Another way is that part of the stored energy is used to power  circuits and the rest is expected to be used in the next slot. However, such an assumption makes our considered problem more complex.}}.} In other words, there will be no energy stored in  node $A$ at the end of each  slot.
An entire slot is less than the coherence interval, which is normalized to 1 without loss of generality. There are two states, sleep state $\tau_s$ and active state $\tau_a$, in one slot. {\color{black}Let $h_0$, $h_1$, $h_2$ denote the channel gains of $S-A$ link, the $A-B$  link, and $S-B$ link, respectively.}  Each link is assumed to undergo independent identically distributed quasi-static fading and  to be reciprocal. {\color{black}Motivated by the recent works \cite{7876867, 7981380,8636518, 8093703, 8468064, 7937935, 8424210,8170328,8490651,8302460,8527670} in this filed, we assume  perfect CSI is available to obtain an EE upper bound of a WP-BackCom.   The way to obtain  $h_0$ and $h_1$ can be found in \cite{8527670} and  $h_2$ can be estimated by the conventional traditional pilot-based method. Relaxing this assumption into  imperfect CSI makes our considered network more realistic, which can be studied in our future work.}


For each slot,  node $A$ leverages the RF signals  $x(n)$ {\small{$\left( {\mathbb{E}\left[ {{{\left| {x\left( n \right)} \right|}^2}} \right] = 1} \right)$}} from $S$ to realize information transmission and energy harvesting for circuit operation.   Node $A$ firstly operates in the sleep state to harvest energy from received RF signals and the harvested energy in this state is calculated as ${E_{\rm{sleep}}^h} = \eta {P_{\rm{0}}}{h_0}{\tau _s}$,
where $P_0$ and $\eta$ are the transmit power of $S$ and the energy harvesting (EH) efficiency coefficient\footnote{In this work, we   assume a linear EH model   for analytical tractability, which is the same as \cite{ 8490651,7876867,7981380,8093703,8468064,7937935, 8424210,8170328,8302460}.} , respectively.
Here we ignore the harvested energy from the noise  since the thermal noise power of the passive node $A$    is much smaller than the received signals $x(n)$ \cite{7981380, 7937935,8468064, 8424210, 8093703}.
In the active state,  part of  the received RF signal, {\small{$\sqrt {\beta P_0{h_0}} x\left( n \right)$}}, is employed by the vehicle for modulating and backscattering the information of  node $A$  and the rest, {\small{$\sqrt {(1-\beta) P_0{h_0}} x\left( n \right)$}}, is flown into the energy harvester. For convenience, we refer {\small{$0\!<\!\beta\!\le\!1$}} as the reflection coefficient \cite{8468064,7876867,8093703,8424210}. Thus, the harvested energy in this state and the backscattered signals are written as {\small{${E_{\rm{active}}^h} = \eta \left( {1 - \beta } \right){P_0}{h_0}{\tau _a}$}} and {\small{$\tilde x\left( n \right) = \sqrt {\beta {P_0}{h_0}} x\left( n \right)c\left( n \right)$}}, respectively, where $c(n)$ is  node $A$'s signal satisfying {\small{$\mathbb{E}\left[ {{{\left| {c\left( n \right)} \right|}^2}} \right] = 1$}} \cite{8424210}. The received signal at node $B$ is given by {\small{$y\left( n \right) = \sqrt {{h_1}} \tilde x\left( n \right) + \sqrt {P_0{h_2}} x\left( n \right) + w\left( n \right)$}},
where $w\left( n \right)$ is the additive white Gaussian noise with variance $\sigma^2$ at the receiver. {\color{black}In our work, the energy source $S$ only serves as a RF source and hence the transmitted RF signal $x\left( n \right)$ can be  a predefined pattern that node $B$ knows.
 Once the CSI is obtained by node $B$, $\sqrt {P_0{h_2}} x\left( n \right)$  can be subtracted in $y(n)$ by using existing digital or analog cancellation techniques. For this reason,  the received signal-to-noise ratio (SNR) is calculated as {\small{$\gamma  = \frac{{\beta{P_0}{h_0}{h_1}}}{{{\sigma ^2}}}$}} after applying successive interference cancellation (SIC) at the node $B$ \cite{8424210,8468064,8093703,8302460}}.  Then the  throughput\footnote{Indeed, it is difficult to evaluate the exact throughput of a backscatter communication  as the distribution of  $\tilde x\left( n \right)$ is indeterminate \cite{7551180}. A common
 way is to assume $\tilde x\left( n \right)$ as a complex Gaussian  and use Shannon equation to approximate the maximum achievable throughput  \cite{8424210,8468064,8093703,8490651, 8302460}.} is
{\small{$R = {\tau _a}{\log _2}\left( {1 + \beta {P_0}\lambda } \right)$}},
where $\lambda={\frac{{{h_0}{h_1}}}{{{\sigma ^2}}}}$.

The total energy consumption consists of two parts: the energy consumed  in the dedicated energy RF source and the energy consumed  in node $B$. Therefore,  the total energy consumption of the whole system is written as
$E_{{\rm{total}}}^{\rm{c}} = \frac{{{P_0}{\tau _s}}}{\xi }+P_{sc}\tau_s +\frac{{{P_0}{\tau _a}}}{\xi } + {P_{sc}}{\tau _a} +  {P_{rc}}{\tau _a}$,
where $\xi  \in \left( {0,1} \right]$ is the   power amplifier  efficiency; $P_{sc}$ and  $P_{rc}$  are the constant circuit powers consumed by $S$ and node $B$, respectively.
 Note that the constant circuit power of node $A$, denoted by  $P_{tc}$, is not included in $E_{{\rm{total}}}^{\rm{c}}$ since  node $A$ is powered by the harvested energy, which has been included in the energy consumption of the energy RF source $S$. 
\section{Energy-efficient Resource Allocation}
\subsection{Problem Formulation}
{\color{black}In this subsection, we formulate an optimization problem to maximize the achievable EE by optimizing the time for sleep  and active states, reflection coefficient, and transmit power of $S$.} The  EE $q$  is defined as the ratio of achievable throughput to the total energy consumption \cite{6661329}, given as
{\small{$q = \frac{{{\tau _a}\log_2 \left( {1 + \beta {P_0}\lambda } \right)}}{{\frac{{{P_0}}}{\xi }\left( {{\tau _a} + {\tau _s}} \right) + {P_{rc}}{\tau _a} + {P_{sc}}\left( {{\tau _a} + {\tau _s}} \right)}}
=\frac{{\log_2 \left( {1 + \beta {P_0}\lambda } \right)}}{{\frac{{{P_0}}}{\xi }\left( {{\rm{1}} + \frac{{{\tau _s}}}{{{\tau _a}}}} \right) + {P_{rc}} + {P_{sc}}\left( {{\rm{1}} + \frac{{{\tau _s}}}{{{\tau _a}}}} \right)}}$}}.
Thus, the optimization problem is formulated in the following.
{\color{black}
\begin{small}
\begin{align}\label{A}
\begin{array}{l}
\mathbf{P}_1:\mathop {\max }\limits_{{P_0},{\tau_s},{\tau_a},\beta } q\\
\;\;{\rm{s.t}}.\;{\rm{C}}1:0 < \beta  \le 1,\;{\rm{C2}}:{\tau _a} + {\tau _s} = 1,\\
\;\;\;\;\;\;\;\;{\rm{C}}3:0<{P_0} \le {P_{\max }},\;{\rm{C4}}:{\tau _a>0}, {\tau _s} \ge 0,\\
\;\;\;\;\;\;\;\;{\rm{C5}}:{P_{tc}}{\tau_a} \le {E_{\rm{sleep}}^h} + {E_{\rm{active}}^h}.
\end{array}
\end{align}
\end{small}}In $\mathbf{P}_1$,  $\rm{C}3$  constrains the maximum transmit power of $S$. $\rm{C}5$ guarantees that the total energy consumed by  node $A$ does not exceed the total harvested energy \cite{8093703}. {\color{black}Note that the WP-BackCom is  different from the conventional relaying and the main differences can be found in \cite{8468064}}. These differences make the formulated EE problem noticeably different from that of the conventional relaying.

Obviously, $\mathbf{P}_1$  is a non-convex problem due to the non-convex objective function and the non-convex constraint $\rm{C}5$. In general, there is no standard algorithm to solve non-convex optimization problems efficiently. We propose an iterative algorithm to solve $\mathbf{P}_1$ in what follows.
\vspace{-5pt}
\subsection{Solution}
The problem $\mathbf{P}_1$  is a non-linear fractional programming problem and hence this can be solved by developing an efficient Dinkelbach-based iterative algorithm.
To this end, Lemma 1 is provided to transfer $\mathbf{P}_1$  to a tractable problem.

\textbf{Lemma 1.} The optimal solution of $\mathbf{P}_1$  can be obtained if and only if
{\small{$\mathop {\max }\limits_{{P_0},{\tau _s},{\tau _a},\beta } \log_2 \left( {1 + \beta {P_0}\lambda } \right) - q\left( {\left( {\frac{{{P_0}}}{\xi } + {{P_{sc}}}} \right)\left( {{\rm{1}} + \frac{{{\tau _s}}}{{{\tau _a}}}} \right) + {P_{rc}}} \right)
 =\log_2 \left( {1 + {\beta ^*}P_0^*\lambda } \right) - q\left( {\left( {\frac{{P_0^*}}{\xi } + {{P_{sc}}}} \right)\left( {{\rm{1}} + \frac{{\tau _s^*}}{{\tau _a^*}}} \right) + {P_{rc}}} \right)
 = 0$}} holds,
{\normalsize{where}} $*$ denotes the optimal solution corresponding to the optimization variables. This Lemma can be proven readily  from the generalized fractional programming theory \cite{6661329}.

Based on Lemma 1, the original  problem $\mathbf{P}_1$ can be solved by solving the following problem $\mathbf{P}_2$.
\begin{small}
\begin{align} \label{D}\notag
\mathbf{P}_2: &\mathop {\max }\limits_{{P_0},{\tau _s},{\tau _a},\beta }\! \log_2 \left( {1 \!+\! \beta {P_0}\lambda } \right) \!- \!q\left(\! {\left(\! {\frac{{{P_0}}}{\xi } + {{P_{sc}}}} \right)\left( {{\rm{1}} + \frac{{{\tau _s}}}{{{\tau _a}}}}\! \right) + {P_{rc}}} \!\right)\\
&\;\;\;\;\;{\rm{s}}{\rm{.t}}{\rm{.}}\;\;\rm{C}1-C5.
\end{align}
\end{small}Even though the problem is more tractable, there are coupling relationships among different optimization variables.
 Accordingly, the problem $\mathbf{P}_2$ is still non-convex.  In order to solve it,  we first present the following Proposition.

\textbf{Proposition 1.} For any given system parameters and optimization variables, the optimal reflection coefficient ${\beta ^*}$ of $\mathbf{P}_2$  is calculated as {\small{${\beta ^*} = \max \left\{ {0,\min \left\{ {1 + \frac{{{\tau _s}}}{{{\tau _a}}} - \frac{{P_{tc}}}{{\eta {P_0}{h_0}}},1} \right\}} \right\}$}}.

\emph{Proof.}  Obviously, the objective function of $\mathbf{P}_2$  increases with the increase of $\beta$. On the other hand, through some simple mathematical calculations, the constraint $\rm{C}5$ is equivalent to the following inequality, which is $\beta  \le 1 + \frac{{{\tau _s}}}{{{\tau _a}}} - \frac{{P_{tc}}}{{\eta {P_0}{h_0}}}$. Combining with  $\rm{C}1$, the Proposition 1 can be proven.

\emph{Remark 1.} The proposed Proposition 1 serves two  purposes. Firstly, we provide a closed-form expression for the optimal reflection coefficient and hence obtain the optimal reflection coefficient using this expression instead of other iterative algorithms.  The second purpose is to  obtain insightful understandings on  the optimal reflection coefficient. For example, when $0 \le 1 + \frac{{{\tau _s}}}{{{\tau _a}}} - \frac{{P_{tc}}}{{\eta {P_0}{h_0}}} < 1$ holds, the optimal reflection coefficient increases with the increase of $\tau_s$, and more power of the  (or even all the) received signals in the active state  will be used to backscatter, indicating that a higher EE could be achieved; when $1 + \frac{{{\tau _s}}}{{{\tau _a}}} - \frac{{P_{tc}}}{{\eta {P_0}{h_0}}} \ge1$ is satisfied, i.e., the harvested energy during sleep state is sufficient
to cover the energy consumed by circuits, the transmitter
backscatters all the received signals during the active state and assigns more time for the active
state and less time for the sleep state for EE maximization.


Based on Proposition 1,  $\mathbf{P}_2$ is rewritten as
\begin{small}
 \begin{align}\label{C}
\begin{array}{l}
\mathbf{P}_3: \mathop {\max }\limits_{{P_0},{\tau _s},{\tau _a}} \log_2 \left( {k + {P_0}\lambda \left( {1 + \frac{{{\tau _s}}}{{{\tau _a}}}} \right)} \right)\\
\;\;\;\;\;\;\;\;\; - q\left( {\left( {\frac{{{P_0}}}{\xi } + {P_{sc}}} \right)\left( {{\rm{1}} + \frac{{{\tau _s}}}{{{\tau _a}}}} \right) + {P_{rc}}} \right)\\
{\rm{s}}.{\rm{t}}.\;\;\;\;\;{\rm{C2}}-{\rm{C4}}, \;{\rm{C6}}:\; 0 < 1 + \frac{{{\tau _s}}}{{{\tau _a}}} - \frac{{P_{tc}}}{{\eta {P_0}{h_0}}} \le 1,
\end{array}
\end{align}
\end{small}where $k=1-{\frac{{\lambda P_{tc}}}{{\eta {h_0}}}}$, and  $\rm{C}6$ is derived from $\rm{C}1$ and Proposition 1.
Observe that the  problem $\mathbf{P}_3$ has less optimization variables and more tractable compared with the original problem $\mathbf{P}_2$. However, the  problem $\mathbf{P}_3$ is still non-convex due to the existence of coupling in the objection function and the constraint $\rm{C}6$. {\color{black} To cope with it, we introduce two  auxiliary variables: $z = {P_0}\left( {1 + {{{\tau _s}} \mathord{\left/
 {\vphantom {{{\tau _s}} {{(1-\tau _s)}}}} \right.
 \kern-\nulldelimiterspace} {{(1-\tau _s)}}}} \right)$ and $t={1 + \frac{{{\tau _s}}}{{{1-\tau _s}}}}$.} Based on these two  auxiliary variables, the   problem $\mathbf{P}_3$  is equivalent to the following problem, given by
{\color{black}\begin{small}
\begin{align}\label{E}
\begin{array}{*{20}{l}}
\mathbf{P}_4: {\mathop {\max }\limits_{z,t} \log_2 \left( {k + \lambda z} \right) - q\left( {\frac{z}{\xi } + {{P_{sc}}}t + {P_{rc}}} \right)}\\
{{\rm{s}}.{\rm{t}}.\;\;\;}\;{\rm{C}}7:0 < z \le {P_{\max }}t,\\
{\;\;\;\;\;\;\;\;\;{\rm{C}}8:t \ge 1},\;{\rm{C}}9:0 < t\left( {1 - {{P_{tc}} \mathord{\left/
 {\vphantom {{P_{tc}} {\eta z{h_0}}}} \right.
 \kern-\nulldelimiterspace} {\eta z{h_0}}}} \right) \le 1,
\end{array}
\end{align}
\end{small}where the constraints $\rm{C}8$ and $\rm{C}9$ are derived from the constraints $\rm{C}4$ and $\rm{C}6$, respectively.}


The problem $\mathbf{P}_4$  is still non-convex due to the non-convex constraint $\rm{C}9$, while we note that the objective function increases with the decrease of $t$ and the feasible region of $t$ is ${\rm{max}} \{ 1,{z \mathord{\left/
 {\vphantom {z {{P_{\max }}}}} \right.
 \kern-\nulldelimiterspace} {{P_{\max }}}}\}  \le t \le \frac{1}{{1 - {{P_{tc}} \mathord{\left/
 {\vphantom {{P_{tc}} {\eta z{h_0}}}} \right.
 \kern-\nulldelimiterspace} {\eta z{h_0}}}}}$. Based on this observation, we show that the problem  $\mathbf{P}_4$  is equivalent to the following two optimization
problems $\mathbf{P}_5$ and $\mathbf{P}_6$.
\vspace*{-4pt}
{\color{black}
\begin{small}\begin{align}\label{F}
\begin{array}{*{20}{l}}
\mathbf{P}_5: {\mathop {\max }\limits_{z } \log_2 \left( {k + \lambda z} \right) - q\left( {\frac{z}{\xi } + {P_{sc}} + {P_{rc}}} \right)}\\
{{\rm{s}}.{\rm{t}}.\;\;\;{\rm{C10:}}\;0 < z \le {P_{\max }}},
{\;{\rm{C11}}:\;\eta z{h_0} - {P_{tc}} > 0}.
\end{array}
 \end{align}
 \end{small}}
 \vspace{-8pt}
{\color{black}
\begin{small}
\begin{align}\label{G}
\begin{array}{*{20}{l}}
\mathbf{P}_6: {\mathop {\max }\limits_{z} \log_2 \left( {k + \lambda z} \right) - q\left( {\frac{z}{\xi } + \frac{{z{P_{sc}}}}{{{P_{\max }}}} + {P_{rc}}} \right)}\\
{{\rm{s}}.{\rm{t}}.\;\;\;\;{\rm{C}}12:{P_{\max }} < z},
{\;{\rm{C}}13:\frac{{{P_{tc}}}}{{\eta {h_0}}} \le z \le {P_{\max }} + \frac{{{P_{tc}}}}{{\eta {h_0}}}}.
\end{array}
\end{align}
\end{small}}$\mathbf{P}_5$ and $\mathbf{P}_6$  are formulated based on $\frac{z}{{{P_{\max }}}} \le 1$ and $\frac{z}{{{P_{\max }}}} > 1$, respectively. {\color{black}Obviously,  the problem $\mathbf{P}_5$  (or $\mathbf{P}_6$) is convex and  can be solved by  bisection search method.  The computational complexity for $\mathbf{P}_5$  (or $\mathbf{P}_6$)  is {\small{$O(\log(\frac{D_{1}}{\varepsilon}))$}} (or {\small{$O(\log(\frac{D_{2}}{\varepsilon}))$}}), where ${D_1 }$ (or $D_2$) and ${\varepsilon}$ denote  the maximum range of the searching variable and the precision, respectively. If the number of iterations for Algorithm 1 is $K$, the  overall computational complexity of Algorithm 1 is $KO(\log(\frac{D_{1}D_2}{\varepsilon^2}))$, where $D_1={P_{\max }}-\frac{{{P_{tc}}}}{{\eta {h_0}}}$ and $D_2=\min \left\{ {{P_{\max }},\frac{{{P_{tc}}}}{{\eta {h_0}}}} \right\}$.}

\emph{Remark 2.} If $\frac{z}{{{P_{\max }}}} \le 1$ holds, we have {\color{black}$t^*=1$, $\tau_s^*=0$ and $\tau_a^*=1$}, indicating that the harvested energy during the active state is sufficient to power the circuit and that  node $A$ always operates in the active state. In addition,   we derive the closed-form expressions for $z^*$ and $P_0^*$ based on Lagrange duality method and $z^*=P^*_0t^*$. $z^*=P_0^* = \frac{{\ln 2}}{{{u_1} + {q \mathord{\left/
 {\vphantom {q \xi }} \right.
 \kern-\nulldelimiterspace} \xi }}} - \frac{k}{\lambda }$, where $u_1\ge0$ is a Lagrange multiplier.

\begin{figure*}[t]
\begin{minipage}[t]{0.33\linewidth}
\centering
\includegraphics[width=2.25in,height=1.62in]{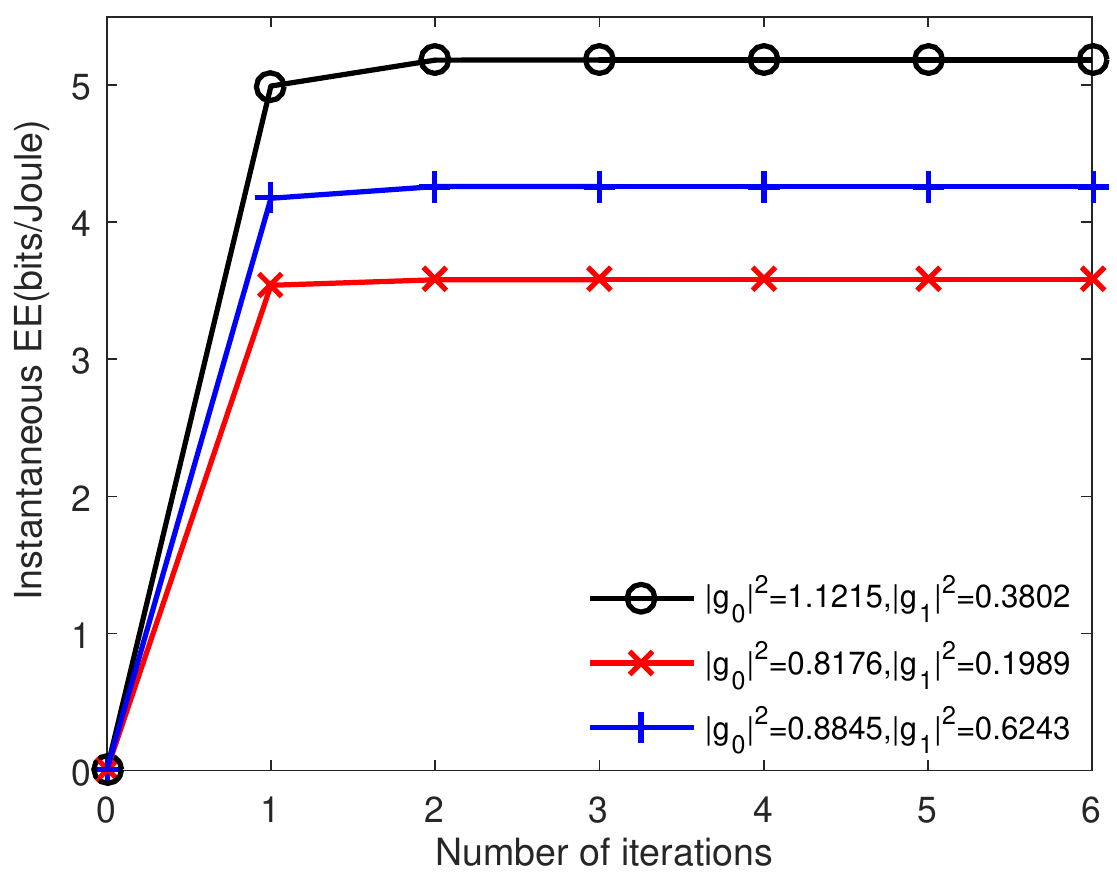}
\centering \caption{The convergence of Algorithm 1.}
\end{minipage}
\begin{minipage}[t]{0.33\linewidth}
\centering
\includegraphics[width=2.25in,height=1.62in]{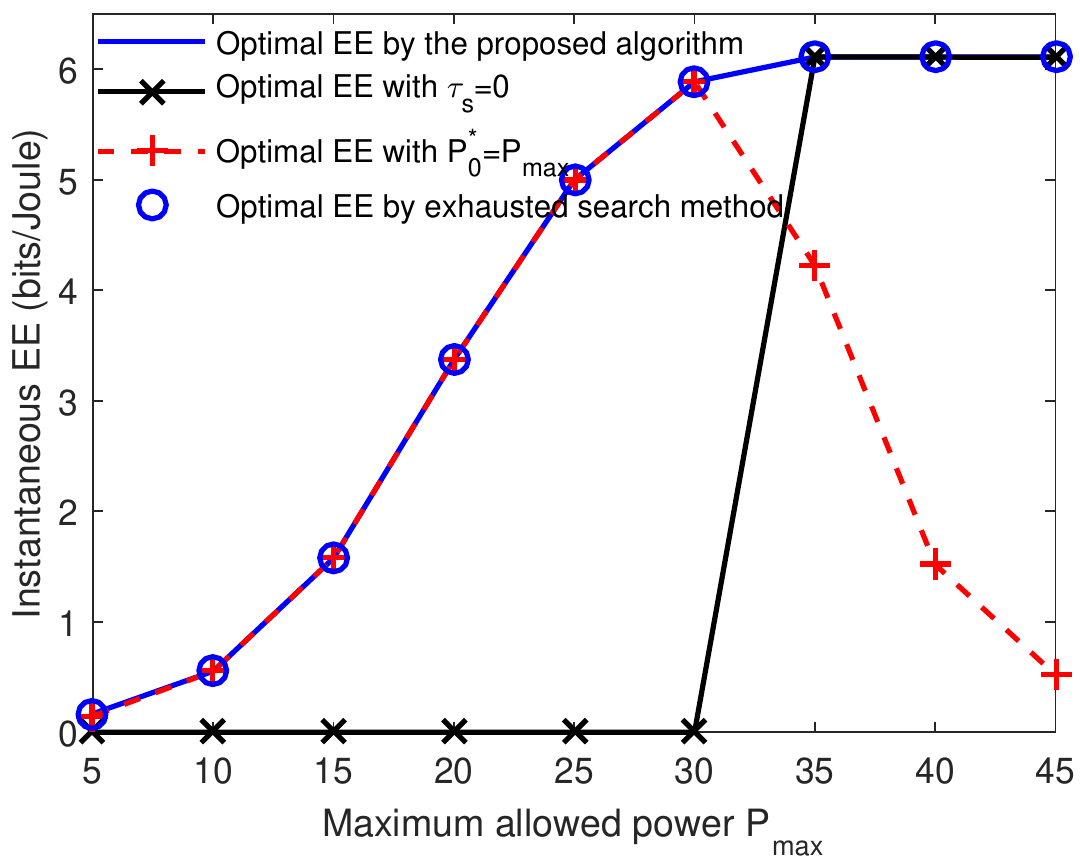}
\centering \caption{{\color{black}An illustration of Remark 4.}}
\end{minipage}
\begin{minipage}[t]{0.33\linewidth}
\centering
\includegraphics[width=2.25in,height=1.62in]{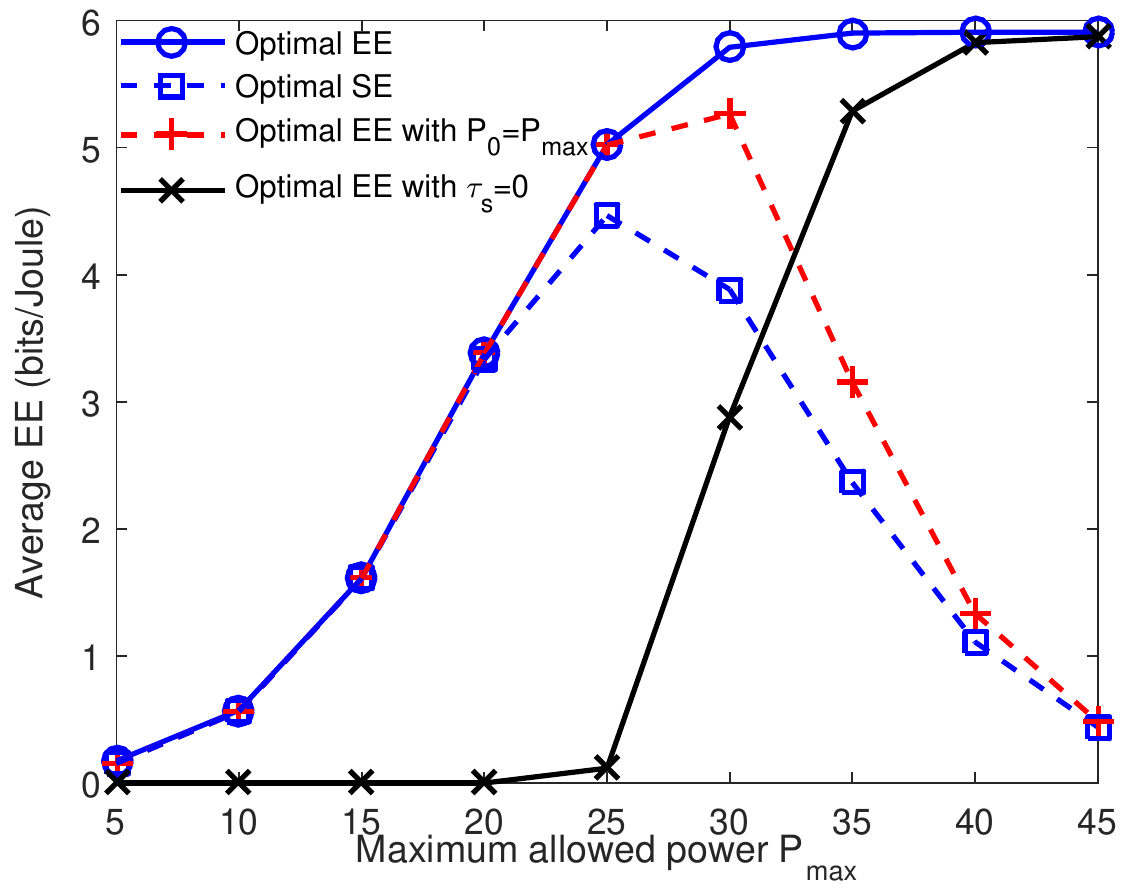}
\centering \caption{EE versus the maximum allowed power.}
\end{minipage}
\vspace{-5pt}
\end{figure*}
\emph{Remark 3.} If $\frac{z}{{{P_{\max }}}} > 1$ is satisfied, {\color{black}we have $t^*=\frac{z^*}{{{P_{\max }}}}$, $\tau_s^*>0$ and $0<\tau_a^*<1$}. Combined with $z^*=P_0^*t^*$, it is not difficult to find that $P_0^*=P_{\rm{max}}$. There are two insights: (i) $\tau_s^*>0$ and $0<\tau_a^*<1$ mean that  \lq sleep-then-active\rq$ $ is  a desirable working mode  for node $A$;
 (ii)  the maximum  EE could be achieved when the $S$  adopts the maximum allowed power. Moreover, we obtain that ${t^*} = \frac{{\ln 2}}{{{{q{P_{\max }}} \mathord{\left/
 {\vphantom {{q{P_{\max }}} \xi }} \right.
 \kern-\nulldelimiterspace} \xi } + {u_2}{P_{\max }} - q{P_{sc}}}} - \frac{k}{\lambda }$ by using Lagrange duality method, where $u_2\ge0$ is a Lagrange multiplier. It can be found that $t^*$ increases with the increase (decrease) of $P_{sc}$ ($P_{\rm{max}}$). This finding and $t^*={1 + \frac{{{\tau^* _s}}}{{{\tau^* _a}}}}$ reveal the relationships among $\frac{{{\tau^* _s}}}{{{\tau^* _a}}}$, $P_{sc}$ and $P_{\rm{max}}$ (or $P^*_0$).

\emph{Remark 4.} It can be drawn from remarks 2 and 3 that the problem $\mathbf{P}_2$  is equivalent to two optimization problems, $\mathbf{P}_5$ and $\mathbf{P}_6$, for two simplified sub-systems, which can be obtained by relaxing one constraint, e.g, $\tau_s^*=0$ or $P^*_0=P_{\rm{max}}$.

Based on $\mathbf{P}_2$$-$$\mathbf{P}_6$, we summarize the Dinkelbach-based iterative algorithm for solving $\mathbf{P}_1$ in Algorithm 1, where $f_1(\cdot)$, $f_2(\cdot)$, $z_1^+$ and $z_2^+$ denote the objective function of $\mathbf{P}_5$, the objective function of $\mathbf{P}_6$, the  optimal solution of  $\mathbf{P}_5$  and the optimal solution of $\mathbf{P}_6$  in each iteration, respectively. In the proposed Algorithm 1, we solve  $\mathbf{P}_5$  and $\mathbf{P}_6$ instead of $\mathbf{P}_4$  with a given $q$ in each iteration and obtain the optimal solution, denoted by $\left( {{z^+},{ t^+}} \right)$, by comparing ${f_1}\left( {z_1^ + } \right)$ with  ${f_2}\left( {z_2^ +  } \right)$. For an error tolerance $\epsilon$, the solution to $\mathbf{P}_4$  is determined when  $\log_2 \left( {k + \lambda {z^ + }} \right) - \frac{{q{z^ + }}}{\xi } - q{P_{sc}}{t ^ + } - q{P_{rc}}<\epsilon$ or $l=L_{\max}$ is satisfied.
\vspace{-5pt}
\begin{algorithm}
\caption{Dinkelbach-based Iterative Algorithm}
\label{alg:A}
\begin{algorithmic}[1]
\STATE {Set the maximum iterations $L_{\max}$, the maximum error tolerance $\epsilon$, the   iteration index $l=0$ and $q=0$.}
\REPEAT
\STATE Solve $\mathbf{P}_5$ and $\mathbf{P}_6$ with a given $q$, and obtain the  optimal solutions $ {{z_1^+}}$ and ${{z_2^+}}$;
\IF {${f_1}\left( {z_1^ + } \right) > {f_2}\left( {z_2^ + } \right)$}
\STATE $z^ +=z_1^ + $ and $t^+=1$
\ELSE
\STATE Set $z^ +=z_2^ +$ and $t^+=\frac{z^+}{{{P_{\max }}}}$
\ENDIF
\IF {$\log_2 \left( {k + \lambda {z^ + }} \right) - \frac{{q{z^ + }}}{\xi } - q{P_{sc}}{t ^ + } - q{P_{rc}}<\epsilon$}
\STATE Set $\textrm{Flag}=1$, $z^*=z^+$, $t^*=t^+$ and return
\ELSE
\STATE Set $\textrm{Flag}=0$, $q = \frac{{\log_2 \left( {k + \lambda {z^ + }} \right)}}{{{z^ + }}}$ and $l=l+1$
\ENDIF
\UNTIL{$\textrm{Flag}=1$ or $l=L_{\max}$}
\STATE Obtain the optimal solution for (\ref{A}) as follows: $P_0^*=\frac{z^*}{t^*}$, $\tau _s^*=1-\frac{1}{t^*}$, $\tau _a^* = 1-\tau _s^*$, $\beta^*=1 + \frac{{\tau _s^*}}{{\tau _a^*}} - \frac{{{P_c}}}{{\eta P_0^*{h_0}}}$.
\end{algorithmic}
\end{algorithm}

\vspace{-12pt}
\section{Simulation Results}
We adopt the distance-dependent path loss model ${h_i} = {\left| {{g_i}} \right|^2}d_i^{-3}$, where ${g_i} \sim \mathbb{CN}\left( {0,1} \right)$ is  the channel coefficient. We set the other parameters as follows: $d_0=10$ m, $d_1=15$ m, $\xi=0.9$, $P_{sc}=100$ mW, $P_{tc}=1$ mW, $P_{rc}=10$ mW, ${\sigma ^2} =  - 100$ dBm, and $\eta=0.6$.


Fig. 2  depicts the  EE of the proposed Algorithm 1 versus the number of iterations  under different channel coefficients. It can be seen that Algorithm 1 converges to the optimal EE after only two iterations.
A concrete example to verify \emph{Remark 4} is presented in Fig. 3, where $|g_0|^2$ and $|g_1|^2$ are set to the unit value  and the step of the maximum allowed power $P_{\rm{max}}$ is $5$ dBm.
{\color{black}It can be seen that the results achieved by the proposed algorithm match the exhaustive search results well and this verify our proposed iterative algorithm.}
It is also shown that the energy-efficient WP-BackCom network  operates   as expected in both modes,  namely the mode $1$ where the dedicated energy RF source adopts the maximum allowed power or the mode $2$ where the transmitter always operates in the active state. Besides, our study shows that the considered network  switches from mode 1 to mode 2 as $P_{\rm{max}}$ increases.

{\color{black}In Fig. 4, we plot the average  EE versus the maximum allowed power $P_{\rm{max}}$ for four schemes, given by (i) optimal EE proposed  in this letter; (ii) optimal SE to maximize  the  throughput  \cite{8093703}; (iii) optimal EE with $P_0=P_{\rm{max}}$ as in \eqref{G}; (iv) optimal EE with $\tau_s=0$ as in \eqref{F}. The average EE of each of the above four schemes  is obtained through $500$ Monte-Carlo simulations. It can be seen that as $P_{\rm{max}}$ increases, the average EE in both  the first case and the fourth case first increases and then remains unchanged while the average EE of the optimal SE scheme and optimal EE with $P_0=P_{\rm{max}}$ first increase and then strictly decrease due to its greedy usage of power. By comparisons, we can see that our proposed scheme always achieves the highest EE among four schemes since the proposed scheme can utilize the power and time resources more efficiently. Besides,  with a small $P_{\rm{max}}$, the average EE of the first case is the same as that of the third case due to the fact that when the transmit power is small  the considered network switches to mode $1$ to harvest more energy for circuit operation. Similarly, for a larger $P_{\rm{max}}$, the average EE of the first case is the same as that of the fourth case since with a large transmit power, the harvested energy during the active state is sufficient to power the circuit.}


\section{Conclusions}
{\color{black}In this letter, we proposed an energy-efficient resource allocation scheme with a Dinkelbach-based iterative algorithm
to obtain the optimal  time allocation, the optimal reflection coefficient, and the optimal transmit power of the dedicated RF  energy source in a WP-BackCom network.} We verified the fast convergence of the proposed iterative algorithm. It was also shown that, for each iteration, the  energy-efficient WP-BackCom network can  function as either  the  network in which the transmitter always operates in the active state, or the network in which the dedicated energy RF source adopts the maximum allowed power.

\ifCLASSOPTIONcaptionsoff
  \newpage
\fi
\bibliographystyle{IEEEtran}
\vspace{-8pt}
\bibliography{ref}
\end{document}